\documentclass[12pt, fleqn]{article}
\usepackage[cp1251]{inputenc}
\usepackage{latexsym,amsfonts,amssymb}
\usepackage{graphicx}

\newtheorem{theo}{Theorem}
\newcommand{\bt}{\begin{theo}}
\newcommand{\et}{\end{theo}}
\newcommand{\bd}{\begin{displaymath}}
\newcommand{\ed}{\end{displaymath}}

\newcommand{\lf}{\left}
\newcommand{\rg}{\right}

\newcommand{\be} {\begin{equation}}
\newcommand{\ee} {\end{equation}}
\newcommand{\ba} {\begin{array}}
\newcommand{\ea} {\end{array}}

\newcommand{\al} {\alpha}
\newcommand{\lbd} {\lambda}

\begin{document}
 \begin{center}
 {\Large \bf New conditional symmetries \\
 and exact solutions of
 nonlinear \\
 reaction-diffusion-convection equations. II}\\
 %%%\medskip\\
{\bf Roman Cherniha}
%%\footnote{\small e-mail: cherniha@imath.kiev.ua}
 {\bf and  Oleksii Pliukhin}
 %%\footnote{\small e-mail:pliukhin@imath.kiev.ua }
 \\
{\it  Institute of Mathematics, Ukrainian National Academy
of Sciences,\\
 Tereshchenkivs'ka Street 3, Kyiv 01601, Ukraine}\\
 \medskip
 E-mail: cherniha@imath.kiev.ua and pliukhin@imath.kiev.ua
\end{center}

\begin{abstract}
In the first part of this paper \cite{ch-pl-06ar}, a complete description of $Q$-conditional symmetries  for two
classes of reaction-diffusion-convection equations with power
diffusivities is derived. It was shown that  all the known    results  for reaction-diffusion equations
with power diffusivities follow as particular cases from those obtained in \cite{ch-pl-06ar} but not vise versa.
In the second part  the symmetries obtained in  are successfully 
applied for constructing exact solutions 
of the relevant equations. In the particular case, new  exact solutions of nonlinear 
 reaction-diffusion-convection (RDC) equations
arising in application and their natural generalizations are found.

\end{abstract}

\begin{center}
{\bf 1.Introduction.}
\end{center}

This paper is a natural continuation  of  \cite{ch-pl-06ar}.
We apply step by step the  $Q$-conditional symmetries obtained to construct 
exact solutions of the relevant nonlinear RDC equations, including   the Murray equation with the fast and slow
diffusions and the  Fitzhugh-Nagumo  equation with the fast diffusion and convection. 

It is well-known (see e.g. examples in \cite{ch-96,ch98}) that new
non-Lie ans\"atze don't guarantee construction of new exact
solutions. It turns out the relevant  exact solutions may be also
obtainable by the standard Lie machinery if the given equation
admits a non-trivial Lie symmetry. Here  we construct exact
solutions using  the $Q$-conditional symmetry operators  
and show that they are so called non-Lie solutions, i.e. cannot be
obtained using Lie symmetry operators.
 As  it follows from the proofs presented in section 3 \cite{ch-pl-06ar}, the $Q$-conditional symmetry 
 operators have essentially simpler structure if one
 uses the substitution
\be\label {12} V= \cases{
\medskip
 {U^{m+1},\ m\neq-1,}
\cr
 {\ln U,\ m=-1.}}
\ee
 So we will firstly find  exact solutions of equations of the form
  \[ V_{xx}=V^{n}V_{t}-\lambda V_{x}+F(V),
 \] \[ \label {14} V_{xx}=\exp(V)V_{t}-\lambda
V_{x}+F(V), \]  \[ \label{60} V_{xx}=V^nV_{t}-\lambda V^{n+1}
V_x+F(V),\]
 \[ \label {61} V_{xx}=\exp(V)V_{t}-\lambda \exp(V)
V_{x}+F(V).\]
 and afterwards use (\ref{12}) to obtain those  of the RDC equations 
  \be \label {1}
 {U_t}= \ [U ^{m} U_{x}]_{x}+  \lambda U^{m}
 U_{x}+C(U),\ee
 \be \label{1b} U_t=[U^mU_x]_x+\lbd
U^{m+1}U_x+C(U).\ee

\begin{center}
\textbf{2. Exact solutions of nonlinear RDC equations}
\end{center}

We start from the case  $(i)$ of Theorem 1 \cite{ch-pl-06ar}. Equation
\[ \ \ \ U_t= \ [U ^{m} U_{x}]_{x}+  \lambda
U^{m}U_{x}+(\lambda_1 U^{m+1} + \lambda_2)( U^{-m}-\lambda_3),\
m\neq-1, \lambda_2 \not=0 \] and operator \[ Q= \
\partial_t\ +( \lambda_1 U + \lambda_2 U^{-m})\partial_U; \]  are transformed by the
substitution (\ref{12}) to the forms \be\label{2-2}
V_{xx}=V^{n}V_{t}-\lambda V_{x}+(\lambda_1^{*}
V+\lambda_2^{*})(\lambda_{3}-V^{n}),\ee and \[
Q=\partial_{t}+(\lambda_1^{*} V+\lambda_2^{*})\partial_V,\] where
$\lambda_i^*=\lambda_i(m+1),\ i=1,2$.
 The relevant ansatz  is constructed using the standard
procedure, i.e. we solve the linear equation   $Q(V)=0$.
Since its general solution depends on  $\lambda_1^{*}$ two ans\"atze are obtained:
 \be\label {2-4}
V=\cases{ \lambda_2^{*}t+\varphi(x), \quad \ \lambda_1^{*}=0,\cr
\varphi(x)e^{\lambda_1^{*}t}-{\lambda_2^{*} \over \lambda_1^{*}}, \,
\lambda_1^{*}\neq0 } \ee being $\varphi(x)$ an unknown function.
Substituting  (\ref{2-4}) with $\lambda_1^{*}=0$  into (\ref{2-2}),
one arrives at the ordinary differential equation (ODE)
\[\varphi_{xx}+\lambda\varphi_x-\lambda_2^{*}\lambda_3=0,\
\] with the general solution
\[\varphi=c_1+c_2e^{-\lambda x}+{\lambda_2^{*}\lambda_3
\over \lambda}x.\] Hereafter $c_1$ and $c_2$ are arbitrary
constants. Hence  equation (\ref{2-2}) with $\lambda_1^{*}=0$
possesses the exact solution
\[V=\lambda_2^{*}t+c_1+c_2e^{-\lambda
x}+{\lambda_2^{*}\lambda_3 \over \lambda}x.\]

Using substitution   (\ref{12}), we obtain  the exact solution
\be\label {2-5}U=\lf[\lambda_2(m+1)t+c_1+c_2e^{-\lambda
x}+{\lambda_2\lambda_3 (m+1) \over \lambda}x\rg]^{1\over m+1}\ee of
the RDC equation with power nonlinearities \be\label {2-6} U_t=[U^m
U_x]_x+\lambda U^m U_x+\lambda_2 U^{-m}-\lambda_2\lambda_3,\
m\ne-1.\ee Using the result of \cite{ch-se-98, ch-se-2006} one
establishes that equation (\ref{2-6}) (with arbitrary coefficients)
is invariant only under two-dimensional algebra with the basic
operators
 $\partial_t$ and
$\partial_x$. So,  $U=U(c_3x+c_4t),\ c_3,\ c_4\in\mathbb{R}$ is the
most general form of  solutions that are obtainable by Lie
machinery. Obviously, the exact solution  presented above has
different structure and cannot be reduced to this form
 therefore  it is a non-Lie solution. Note this solution is  the Lie solution if one additionally
 sets $c_2=0$. In quite similar way it can be shown that all solutions obtained below are also  non-Lie solutions
 and may be reduced to Lie solution only under additional constraints.

Substituting  (\ref{2-4}) with $\lambda_1^{*}\not=0$  into
(\ref{2-2}), one again obtains a linear second-order ODE, which are
integrable in terms of different elementary functions depending on
$\delta=\lambda^2+4\lambda_1^*\lambda_3=\lambda^2+4\lambda_1(m+1)\lambda_3$.
%%After the relevant straightforward calculations
Dealing in quite similar way to the case $\lambda_1^{*}=0$, we
finally obtain three exact solutions
\be\label {2-7}
U=\lf[\exp\lf(\lambda_1(m+1)t-{\lbd\over2}x\rg)\lf( c_1 \exp
\biggr(\frac{\sqrt\delta}{2}x\biggr)+c_2\exp
\biggr(\frac{-\sqrt\delta}{2}x\biggr)\rg)-{\lambda_2 \over
\lambda_1}\rg]^{1 \over {m+1}}, \, \delta>0  \ee

 \[U=\lf[ \exp
\lf(-\frac{\lambda}{2}x+\lambda_1(m+1)t\rg)(c_1+c_2 x)-{\lambda_2
\over \lambda_1}\rg]^{1 \over {m+1}}, \, \, \delta=0 \]

\be\label {2-9} U=\lf[\exp
\lf(-\frac{\lambda}{2}x+\lambda_1(m+1)t\rg)\lf(c_1\cos
\frac{\sqrt{-\delta}}{2}x+c_2\sin
\frac{\sqrt{-\delta}}{2}x\rg)-{\lambda_2 \over \lambda_1}\rg]^{1
\over {m+1}},\, \delta<0 \ee
of the nonlinear RDC equation \be\label
{2-10} U_t= \ [U ^{m} U_{x}]_{x}+  \lambda U^{m}U_{x}+(\lambda_1
U^{m+1} + \lambda_2 )( U^{-m}-\lambda_3),\ m\neq-1.\ee

In the case of the Murray equation with the slow diffusion \be
\label{3*}U_t= \ [U  U_{x}]_{x}+  \lambda UU_{x}+\lambda_1U(1- U),
\ee one notes that $\delta>0$ if $\lambda_1>0$. Hence solution
(\ref{2-7}) takes the form \[
U=\sqrt{\exp\lf(2\lambda_1t-{\lbd\over2}x\rg)\lf( c_1 \exp
\biggr(\frac{\sqrt\delta}{2}x\biggr)+c_2\exp
\biggr(\frac{-\sqrt\delta}{2}x\biggr)\rg)}, \ \delta=
\lambda^2+8\lambda_1. \] This solution unboundedly  grows if $t \to
\infty$ or  $x \to \pm \infty$. More interesting solutions occur in
the case of (\ref{3*}) with the anti-logistic term: \[ U_t= \ [U
U_{x}]_{x}+  \lambda UU_{x}-U(1- U). \]
 Depending on $\delta=\lambda^2-8$  one obtains three types of solutions.
In the  case $m=1,\ \lambda=3,\ \lbd_1=-1,\ \lbd_2=0,\ c_1=-c_2=4$,
solution (\ref{2-7}) is presented on Fig.1. This solution tends to
zero if $t \to \infty$ and satisfies the zero boundary  conditions
for $x=0$ and $x=\infty$.  If $\lambda=2$ then solution (\ref{2-9})
with $m=1$ is valid. In the case  $\lbd_1=-1,\ \lbd_2=0,\ c_1=1,
c_2=0$ this solution is presented on Fig.2.
 We note that the solution is again vanishing if $t \to \infty$, but one  satisfies the zero boundary  conditions
 on the bounded interval $[-\frac{\pi}{2},\frac{\pi}{2}]$.

\begin{figure}[t]
\begin{minipage}[t]{8cm}
\centerline{\includegraphics[width=8cm]{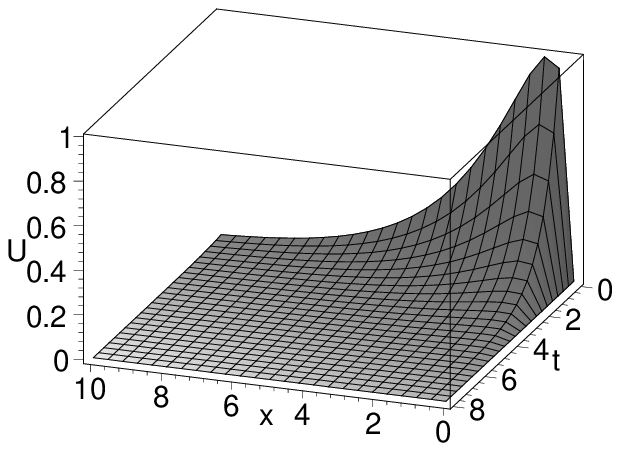}}\center\caption{ Exact
solution (\ref{2-7}) with $m=1,\ \lambda=3,\ \lbd_1=-1,\ \lbd_2=0,\
c_1=-c_2=4$} \label{Fig-1}
\end{minipage}
\hfill
\begin{minipage}[t]{8cm}
\centerline{\includegraphics[width=8cm]{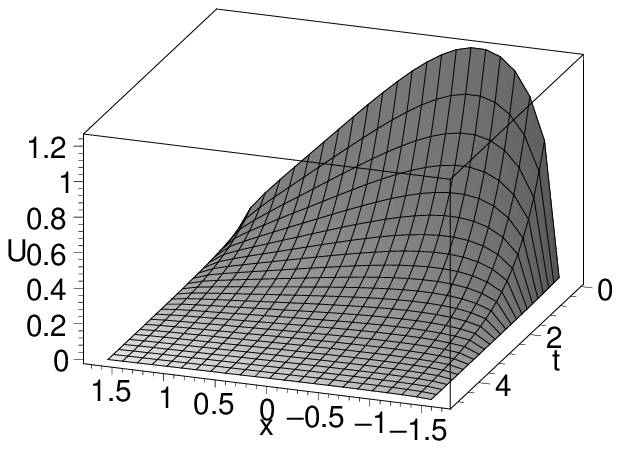}} \center\caption{ Exact
solution (\ref{2-9}) with $m=1,\ \lbd=2,\ \lbd_1=-1,\ \lbd_2=0,\
c_1=1,\ c_2=0$} \label{Fig-2}
\end{minipage}
 \end{figure}

 Consider the Murray equation with the fast
diffusion \be\label{3***}U_t= \ [U^{-2}  U_{x}]_{x}+  \lambda
U^{-2}U_{x}+\lambda_1U(1- U). \ee Since $\delta=\lambda^2>0$
solution (\ref{2-7}) takes the form \be\label {2-7**}
U=\lf[1+\exp(-\lambda_1t)\lf( c_1 + c_2\exp(-\lambda
x)\rg)\rg]^{-1},  \ee which possesses attractive properties.
Assuming $c_1>0$ and $ c_2>0$, one sees that this solution is
positive and bounded for arbitrary $(t,x) \in \mathbb{R^+}\times
\mathbb{R}$. Moreover solution tends either to zero ($\lambda_1<0$)
or to 1 ($\lambda_1>0$)
 if $t \to \infty$. Both values, $U=0$ and $U=1$ are steady-state points of (\ref{3***}).
 Solution (\ref{2-7**}) tends to the steady-state point $U=0$ if $ \lambda x \to -\infty$,
 while $U=[1+c_1\exp(-\lambda_1t)]^{-1}$ if $ \lambda x \to \infty$.
 An example of solution (\ref{2-7**}) is presented on Fig.3 It should be also noted
 that (\ref{2-7**}) with $c_1=0$ is a travelling wave solution with the same structure as
 one for the Murray equation  (see formula (90) in \cite{che-2006}).

Consider the case  $(ii)$ of Theorem 1 \cite{ch-pl-06ar}. Equation
 \be \label {5}\ \ \ U_t= \ [U ^{-1} U_{x}]_{x}+  \lambda
U^{-1}U_{x}+(\lambda_1 \ln U + \lambda_2 )(U - \lambda_3),\
\lambda_1 \not=0,\ee and operator \[ {Q}=\
\partial_t\ +( \lambda_1 \ln U + \lambda_2  )U\partial_U,
\] are transformed by the substitution (\ref{12}) to the forms
\be\label {2-11} V_{xx}=e^{V}V_{t}-\lambda V_{x}+(\lambda_{1}
V+\lambda_{2})(\lambda_{3}-e^{V}),\ee and
 \be\label {2-12}
Q=\partial_{t}+(\lambda_{1} V+\lambda_{2})\partial_V,\ee
respectively. Using operator (\ref{2-12}) we obtain the ansatz
\be\label {2-13} V=\cases{ \lambda_2 t+\varphi(x),\ \quad
\lambda_1=0,\cr
                          \varphi(x) e^{\lambda_1 t}-{\lambda_2 \over
\lambda_1},\, \lambda_1\ne0,}\ee
which has the same structure as (\ref{2-4}).
Substituting  (\ref{2-13})   into (\ref{2-11}),
one again obtains integrable  second-order ODEs and easily constructs  the
relevant  exact solutions of the RDC equation (\ref{5}).
In the case $\lambda_1=0$, the solution is
\be\label {2-14}  U=\exp\lf[\lambda_2 t+c_1+c_2 e^{-\lambda x}+{\lambda_2\lbd_3 \over
\lambda}x\rg],\ee
while the case  $\lambda_1\not=0$ produces three solutions depending on
$\delta=\lambda^2+4\lambda_1 \lambda_3$:

\be\label {2-15}U=\exp\lf[\exp\lf(\lambda_1t-{\lbd\over2}x\rg)\lf(
c_1 \exp \biggr(\frac{\sqrt\delta}{2}x\biggr)+c_2\exp
\biggr(\frac{-\sqrt\delta}{2}x\biggr)\rg)-{\lambda_2 \over
\lambda_1}\rg],\, \delta>0, \ee

\[ U=\exp\lf[\exp
\lf(-\frac{\lambda}{2}x+\lambda_1t\rg)(c_1+c_2 x)-{\lambda_2 \over
\lambda_1}\rg], \, \delta=0 \] and \be\label {2-17} U=\exp\lf[\exp
\lf(-\frac{\lambda}{2}x+\lambda_1t\rg)\lf(c_1\cos
\frac{\sqrt{-\delta}}{2}x+c_2\sin
\frac{\sqrt{-\delta}}{2}x\rg)-{\lambda_2 \over \lambda_1}\rg],\,
\delta<0. \ee Note that properties of solutions
(\ref{2-14})-(\ref{2-17}) depend essentially on values of $c_1$ and
$c_2$. For example, solution (\ref{2-15}) with negative $c_1$ and
$c_2$ tends to zero if $x\to\pm\infty$, while this solution
infinitely increases if those constants are positive. An example of
solution (\ref{2-15}) is presented on Fig.4.

\begin{figure}[t]
\begin{minipage}[t]{8cm}
\centerline{\includegraphics[width=8cm]{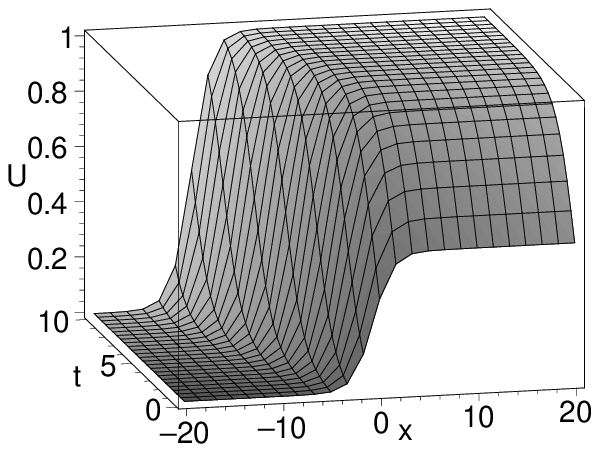}}\center\caption{ Exact
solution (\ref{2-7**}) with $ \lambda=1,\ \lbd_1=1,\ c_1=1,\ c_2=1$}

\end{minipage}
\hfill
\begin{minipage}[t]{8cm}
\centerline{\includegraphics[width=8cm]{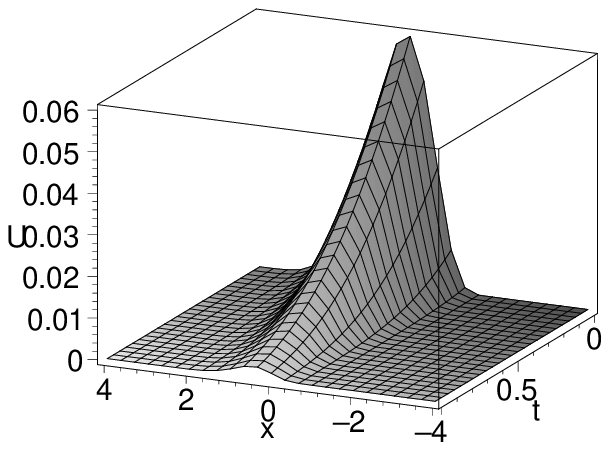}} \center\caption{ Exact
solution (\ref{2-15}) with $\lbd=1,\ \lbd_1=1,\ \lbd_2=1,\
\lbd_3=1,\ c_1=-1,\ c_2=-1$}
\end{minipage}
 \end{figure}
Consider the case  $(iii)$ of Theorem 1 \cite{ch-pl-06ar}. Since we
were unable to solve the overdetermined system
\[ \ba{l}\medskip
2ff_x+f_t+fg=0,\\
\medskip f_{xx}-\lambda f_{x}-2g_{x}-fh=0,\\
\medskip
(g-\frac{\lambda_{1}}{2})(g+2f_{x})+g_{t}=0,\\
\medskip
2gh-\lambda_{1}h+2f_{x}h-\lambda_{2}f_{x}+h_{t}-\lambda g_{x}-g_{xx}=0,\\
\medskip
h^{2}-\frac{\lambda_{2}}{2}h-\lambda_{3}f_{x}+\frac{\lambda_{3}}{2}g-\lambda
h_{x}-h_{xx}=0,\ea\]we used the particular solution producing the
$Q$-conditional operator \[{Q}= \
\partial_t\ + 2h(x)U^\frac{1}{2}\partial_U.\]Application of this operator  leads to a
solution in the implicit form \be\label {2-18}
U=(h(x)t+\varphi(x))^2\ee to the nonlinear RDC equation
\[U_t= \ [U ^{-\frac{1}{2}} U_x]_{x}+
\lambda U^{-\frac{1}{2}}U_{x}+\lambda_2 U^{\frac{1}{2}}\
+\lambda_3.\]
The functions  $h$ and  $\varphi$ arising in (\ref{2-18}) satisfy the ODE system
\[\ba{l}\medskip h_{xx}+\lambda h_x-h^2+\frac{\lambda_2}{2} h=0,\\
\varphi_{xx}+\lambda\varphi_x-h\varphi+\frac{\lambda_2}{2}\varphi+\frac{\lambda_3}{2}=0,\ea\]
which is not integrable. Moreover, there are no any particular solutions of this system
in the known books  \cite {kam, pol-za}. The trivial solution of the first equation
$h=\frac{\lambda_2}{2}$ leads to  a particular
case of  solution (\ref{2-5}).
%%% $(i)$ with $m=-\frac{1}{2},\ \lambda_1=0$ is obtained

Thus, we have constructed all possible exact solutions, which can be
obtained by application of the $Q$-conditional symmetry operators
arising in Theorem 1 \cite{ch-pl-06ar}.

Now we apply the operators arising in Theorem 2 \cite{ch-pl-06ar} to
construct exact solutions. First of all we note that the cases $(i)$
and $(ii)$ only should be considered because for cases $(iv)$ and
$(v)$ the relevant work has been done
 in the recent paper \cite{che-2006}. The case $(iii)$, of course, cannot
 produce any  new results because the Burgers equation is linearizable by the
 Cole-Hopf substitution.

Consider the case  $(i)$ of Theorem 2 \cite{ch-pl-06ar}. The operator
\[ {Q}= \
\partial_t-\lbd U^{m+1}\partial_x +( \lambda_1 U + \lambda_2 U^{-m})\partial_U,\] arising in this case can be successfully  applied to
construct exact solutions in the explicit form.  Omitting rather
trivial computations we present the final result: equation
\[U_t=
\ [U ^m U_x]_x+  \lambda U^{m+1}U_{x}+ \lambda_2 U^{-m},\ m\neq-1;\]
possesses the solution \[ U=\lf[{1\over{\lbd
t+c_1}}\biggr(-x+\lbd_2(m+1)\lf(\lbd\frac{t^2}{2}+c_1t\rg)+c_2\biggl)\rg]^{1\over{m+1}},\]
while \[
U=\lf[{1\over{1+c_1e^{-\lbd_1(m+1)t}}}\lf((m+1)\lf(-{\lbd_1\over\lbd}x+\lbd_2t-{c_1\lbd_2\over
\lbd_1(m+1)}e^{-\lbd_1(m+1)t}\rg)+c_2\rg) \rg]^{1\over{m+1}},\] is
the exact solution of the nonlinear RDC \[U_t= \ [U ^m U_x]_x+
\lambda U^{m+1}U_{x}+\lambda_1 U + \lambda_2 U^{-m},  \] with $\
m\neq-1,\ \lbd_1\ne0$.

 The most cumbersome structure of the conditional symmetry operator
   occurs in case $(ii)$  of Theorem 2 \cite{ch-pl-06ar}.
   As consequence essential difficulties  arise if one applies operator \be\label {53} {Q}=\
\partial_t+\lf(-\lbd U^{1\over2}+{3\lbd_1\over2\lbd}\rg)\partial_x+(\lbd_1 U^{3\over2}+ \lambda_2 U^{1\over
2}+\lbd_3)\partial_U, \ee
   for finding exact solutions.  On the other hand it will be shown
   that  many of the exact solutions obtained possess nice properties.

 Equation \be \label {52}U_t= \ [U ^{-{1\over{2}}} U_x]_x+ \lambda
U^{1\over{2}}U_{x}+(\lbd_1 U^{3\over2}+ \lambda_2 U^{1\over
2}+\lbd_3)\lf({\lbd_1\over {2 \lbd^2}} + U^{1\over2}\rg),\ee and
operator (\ref{53})   by the substitution (\ref{12}) with
 $m=-1/2$ are transformed to the forms
  \be\label{2-21}
V_{xx}=VV_t-\lbd V^2V_x-\frac{\lbd_1^*+3\lbd
V}{3\lbd}\lf({1\over3}\lbd_1^*\lbd V^3+\lbd_2^*V+\lbd_3^*\rg)\ee and
\be\label{2-22} Q=\partial_t+(-\lbd
V+\lbd_1^*)\partial_x+\lf({1\over3}\lbd_1^*\lbd
V^3+\lbd_2^*V+\lbd_3^*\rg)\partial_V,\ee respectively. Hereafter
$\lbd_1^*=\frac{3\lbd_1}{2\lbd}\ne0,\ \lbd_2^*={\lbd_2\over2},\
\lbd_3^*={\lbd_3\over2}$ and $V>0$ is assumed  since (\ref{52})
contains terms  $U^{\frac{1}{2}}$ and $U^{-\frac{1}{2}}$. Instead of
construction of a non-Lie  ansatz using  operator (\ref{2-22}) (in
this case it is a cumbersome procedure), one can use the equation
$Q(V)=0$, i.e. \be\label{2-23}V_t=(\lbd
V-\lbd_1^*)V_x+{1\over3}\lbd\lbd_1^*V^3+\lbd_2^*V+\lbd_3^*,\ee to
eliminate $V_t$  from (\ref{2-21}).  In fact, substituting the
right-hand-side of (\ref{2-23}) into (\ref{2-21}), one arrives at
 \be\label{2-24}
V_{xx}+\lbd_1^*VV_x+{1\over9}{\lbd_1^*}^2V^3+{1\over3\lbd}\lbd_1^*\lbd_2^*+{1\over3\lbd}\lbd_1^*\lbd_3^*=0,\ee
which is the non-linear ODE containing variable $t$ as a parameter.
Equation (\ref{2-24}) is reduced to the form
\be\label{2-25}V_{yy}+3VV_y+V^3+{{3\lbd_2^*}\over{\lbd_1^*\lbd}}V+{{3\lbd_3^*}\over{\lbd_1^*\lbd}}=0\ee
by the simple substitution
 \be\label{2-26}y={\lbd_1^*\over3}x.\ee
 Equation (\ref{2-25}) can be transformed into the linear
third-order ODE \be\label{2-27}W_{yyy}+3pW_y+2q W=0,\ee where
${{\lbd_2^*}\over{\lbd_1^*\lbd}}=p$,
${{3\lbd_3^*}\over{\lbd_1^*\lbd}}=2q$, by the known substitution
\cite{kam}(see item (6.38)) \be\label{2-28}V={W_y\over W}.\ee
According to the classical theory of linear ODE one needs to solve
the algebraic equation \be\label{2-29} k^3+3pk+2q=0,\ee which
corresponds to (\ref{2-27}). Hence four different subcases depending
on the values of $p$ and $ q$ should be separately considered.

Subcase 1.  If  $ p=q=0$  then  $k_{1}=k_2=k_3=0$. The general solution of  (\ref{2-27}) has the form
$W=f+gy+hy^2 $  and we arrive at the expression
\be\label{2-30}V=\frac{g+2hy}{f+gy+hy^2},\ee
giving the general solution of the non-linear ODE (\ref{2-25}).
 Hereafter  $f=f(t),\
g=g(t),\ h=h(t)$ are arbitrary (at the moment) smooth function and at least one of them must be non-zero.
 So (\ref{2-30}) with (\ref{2-26}) generates the general solution of (\ref{2-24}) with $ \lbd_1^*\ne0. $
 Finally, to obtain the general solution of system (\ref{2-21}) and (\ref{2-23}),
 it is sufficiently to substitute (\ref{2-30}) with $y={\lbd_1^*\over3}x$
    into the second
  equation of this system. After the relevant calculations a cumbersome
   expression is obtained, however, one splits
  into separate parts for $x^n, n=0,1,2$ and  we arrive at the ODE system
 \be\label{2-31}\ba{l}\medskip
gh_t-g_th={2\over3}{\lbd_1^*}^2h^2,\\
fh_t-f_th={1\over3}\lbd_1^*h(2\lbd
h+\lbd_1^*g),\medskip\\
fg_t-f_tg={1\over3}\lbd_1^*(2\lbd gh-2\lbd_1^*fh+\lbd_1^*g^2).\ea\ee
System (\ref{2-31}) has the similar  structure to one from
\cite{che-2006} (see formula (60)) and can be solved in a  similar
way.
%%Omitting the relevant details we present only the results
 Substituting the general solution of (\ref{2-31}) into (\ref{2-30}) and using (\ref{2-26}),
 we find the exact solutions
\be\label{2-32}V=\frac{3}{-{\lbd_1^*}^2t+\lbd_1^*x+3c_1}\ee and
\be\label{2-33}V=\frac{{2\lbd_1^*}(x-\lbd_1^*t)+3c_1}{{{\lbd_1^*}^2\over3}
(x-\lbd_1^*t)^2+c_1\lbd_1^*(x-\lbd_1^*t)-2\lbd\lbd_1^*t+c_2\lbd_1^*},\ee
of the equation
 \be\label{2-34} V_{xx}=VV_t-\lbd V^2V_x-{\lbd_1^*\over3}\lbd
V^4-{{\lbd_1^*}^2\over9}V^3.\ee Applying  substitution (\ref{12})
with
 $m=-1/2$ to (\ref{2-32})--(\ref{2-34}) and renaming  parameters, we arrive at the exact solutions
\[U=\lf[\frac{1}{-{3\lbd_1^2\over{4\lbd^2}}t+{\lbd_1\over{2\lbd}}x+c_1}\rg]^2\]
and
\[U=\lf[\frac{2(x-3{\lbd_1\over{2\lbd}}t)+{2\lbd c_1\over\lbd_1}}{{\lbd_1\over{2\lbd}}
(x-{3\lbd_1\over{2\lbd}}t)^2+c_1(x-{3\lbd_1\over{2\lbd}}t)-2\lbd
t+c_2}\rg]^2\]  of the nonlinear RDC equation
\[U_t=[U^{-{1\over2}}U_x]_x+\lbd U^{1\over2}U_x+\lbd_1
U^2+\frac{\lbd_1^2}{2\lbd}U^{3\over2}.\]

Subcase 2.  If  $ p^3=-q^2\ne0$  then  $k_{1}=\al_1=-2\sqrt[3]{q}$ and $k_2=k_3=\al_2=\sqrt[3]{q}$.
The general solution of  (\ref{2-27}) is
\[W=f\exp(\al_1 y)+(g+yh)\exp(\al_2 y), \quad \al_1=-2\al_2\]  so that the expression
\[V=\frac{\al_1f\exp(\al_1
y)+(\al_2g+h(\al_2y+1))\exp(\al_2 y)}{f\exp(\al_1
y)+(g+yh)\exp(\al_2 y)}\] presents the general solution of the
non-linear ODE (\ref{2-25}).
 Dealing in quite similar way to the subcase 1, one easily  obtains the ODE system
\be\label{2-36}\ba{l}\medskip
gh_t-g_th={{\lbd_1^*}\over3}h^2(2\lbd\al_2+\lbd_1^*),\\\medskip
fh_t-f_th=\lbd_1^*\al_2fh(\lbd\al_2-\lbd_1^*),\\
3\al_2(fg_t-f_tg)+fh_t-f_th=\lbd_1^*\al_2f\lf(g({2\over3}\lbd\al_2^2-3\al_2\lbd_1^*-{5\over3})
-h(\lbd\al_2+2\lbd_1^*)\rg)\ea\ee to find the unknown functions
$f(t),\ g(t)$ and $ h=h(t).$ It turns out system (\ref{2-36}) has
the same  structure as one (67) from \cite{che-2006} and its general
solution can be  constructed. Finally, we find the exact solutions
\[U=\lf[\frac{-2c_1\al_2\exp(\beta_0t-\beta_1x)+\al_2c_2}
{c_1\exp(\beta_0t-\beta_1x)+c_2}\rg]^2,\]
and
\[U=\lf[\frac{-2c_1\al_2\exp(\beta_0t-\beta_1x)+c_2\al_2(\beta_2t-\beta_3x+c_3+{1\over\al_2})}
{c_1\exp(\beta_0t-\beta_1x)+c_2(\beta_2t-\beta_3x+c_3)}\rg]^2\]
of the nonlinear RDC equation
\[U_t= \ [U ^{-{1\over{2}}} U_x]_x+ \lambda
U^{1\over{2}}U_{x}+(\lbd_1
U^{3\over2}-3\sqrt[3]{\lbd_1\lbd_3^2\over4} U^{1\over
2}+\lbd_3)\lf({\lbd_1\over {2 \lbd^2}} + U^{1\over2}\rg),\]
where
$\beta_0={3\lbd_1\over2\lbd}\al_2({3\lbd_1\over2\lbd}-\lbd\al_2),\
\beta_1={3\lbd_1\over2\lbd}\al_2,\
\beta_2=-{\lbd_1\over2\lbd}({3\lbd_1\over2\lbd}+2\lbd\al_2),\
\beta_3=-{\lbd_1\over2\lbd},\ \al_2=\sqrt[3]{\lbd_3\over2\lbd_1}\not=0.$

Subcase 3.  If $p^3+q^2<0$  then three roots of (\ref{2-29}) are different
and real. This case is the most cumbersome because the known  Cardano formulae
must be used. Let us set    $k_1=\al_1,$
$k_{2}=\al_2 $ and $\ k_{3}=\al_3,$ where $\al_i, i=1,2,3$ are different real numbers, which
are calculated  by the Cardano formulae
\be\label{2-40}\ba{l}\al_1=-2\sqrt[6]{-p^3}\cos\lf({1\over3}\arctan\lf({\sqrt{-p^3-q^2}\over
q}\rg)\rg),\\
\al_2=2\sqrt[6]{-p^3}\cos\lf({1\over3}\arctan\lf({\sqrt{-p^3-q^2}\over
q}\rg)-{\pi\over3}\rg),\\
\al_3=2\sqrt[6]{-p^3}\cos\lf({1\over3}\arctan\lf({\sqrt{-p^3-q^2}\over
q}\rg)+{\pi\over3}\rg),\ea\ee
if  $q>0$,  by the formulae
\be\label{2-41}\ba{l}\al_1=2\sqrt[6]{-p^3}\cos\lf({1\over3}\arctan\lf({\sqrt{-p^3-q^2}\over
q}\rg)\rg),\\
\al_2=-2\sqrt[6]{-p^3}\cos\lf({1\over3}\arctan\lf({\sqrt{-p^3-q^2}\over
q}\rg)-{\pi\over3}\rg),\\
\al_3=-2\sqrt[6]{-p^3}\cos\lf({1\over3}\arctan\lf({\sqrt{-p^3-q^2}\over
q}\rg)+{\pi\over3}\rg),\ea\ee
if  $q<0$ and by the formulae
\be\label{2-42}\ba{l} \al_1=0,\\
\al_2=\sqrt{-3p},\\ \al_3=-\sqrt{-3p},\ea\ee  if  $q=0$.
 The general solution of  (\ref{2-27}) is
\[W=f\exp(\al_1 y)+g\exp(\al_2 y)+h\exp(\al_3 y),\]  and it leads to the general solution
\be\label{2-43}V=\frac{\al_1f\exp(\al_1 y)+\al_2g\exp(\al_2
y)+\al_3h\exp(\al_3 y)}{f\exp(\al_1 y)+g\exp(\al_2 y)+h\exp(\al_3
y)}\ee
of the non-linear ODE (\ref{2-25}).

Substituting (\ref{2-43}) with $y={\lbd_1^*\over3}x$
    into (\ref{2-23}) and conducting the relevant calculations and splits we again arrive at
    the  ODE system to find the unknown functions $f(t),\
g(t)$ and $ h=h(t).$ This system has the form
\be\label{2-44}\ba{l}\medskip
f_tg-fg_t=-{\lbd_1^*\over3}fg(\lbd(\al_1^2-\al_2^2)+\lbd_2(\al_1-\al_2)),\\\medskip
f_th-fh_t=-{\lbd_1^*\over3}fh(\lbd(\al_1^2-\al_3^2)+\lbd_2(\al_1-\al_3)),\\
g_th-gh_t=-{\lbd_1^*\over3}gh(\lbd(\al_2^2-\al_3^2)+\lbd_2(\al_2-\al_3))\ea\ee
and is fully  integrable \cite{che-2006}. Its general solution leads
to the exact solution \be\label{2-44*} U=\lf[\frac{\al_1
c_1\exp(\beta_1t+\gamma_1 x)+\al_2 c_2\exp(\beta_2t+\gamma_2x)+\al_3
c_3\exp(\beta_3t+\gamma_3 x)}{c_1\exp(\beta_1t+\gamma_1
x)+c_2\exp(\beta_2t+\gamma_2 x)+c_3\exp(\beta_3t+\gamma_3
x)}\rg]^2\ee
\begin{figure}[t]
\begin{minipage}[t]{8cm}
\centerline{\includegraphics[width=8cm]{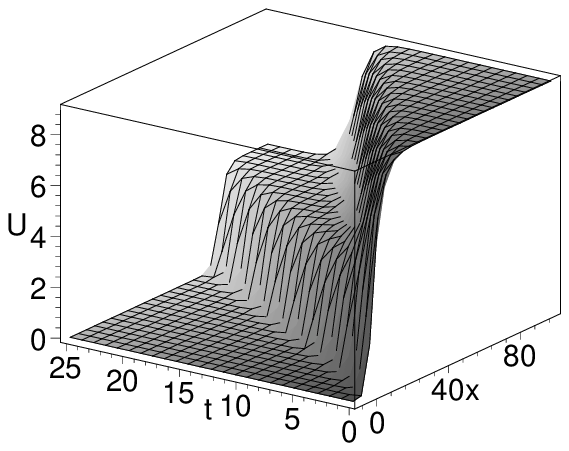}}\center\caption{ Exact
solution (\ref{2-44*}) with $\lambda=1,\ \lbd_1=0.5,\ \
c_1=c_2=c_3=1,\ \al_1=0.1,\ \al_2=2,\ \al_3=3$} \label{Fig-1}
\end{minipage}
\hfill
\begin{minipage}[t]{8cm}
\centerline{\includegraphics[width=8cm]{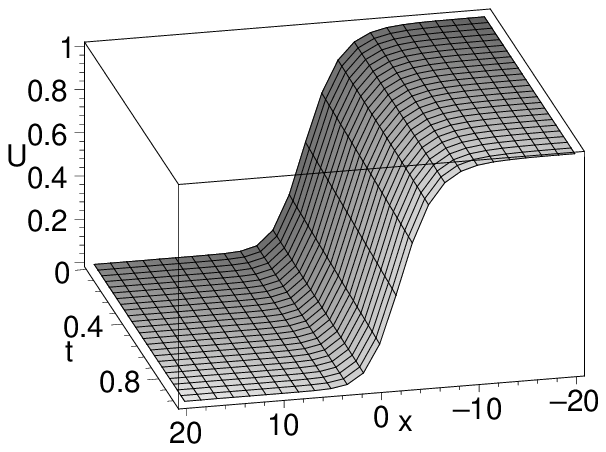}} \center\caption{ Exact
solution (\ref{2-44**}) with $\lbd=1,\ \lbd_2=1,\  c_1=1,\ c_2=1,\
c_3=1,\ \al_2=1,\ \al_3=-1$} \label{Fig-2}
\end{minipage}
 \end{figure}of the nonlinear RDC equation  (\ref{52}) with $\lbd_1\not=0$.
%\[U_t= \ [U^{-{1\over{2}}} U_x]_x+ \lambda U^{1\over{2}}U_{x}+(\lbd_1
%U^{3\over2}+\lbd_2 U^{1\over 2}+\lbd_3)\lf({\lbd_1\over {2 \lbd^2}}+ U^{1\over2}\rg),\]
Here
$\beta_i=-{\lbd_1\over2\lbd}(\lbd\al_i^2+{3\lbd_1\over2\lbd}\al_i),\
\gamma_i={\lbd_1\over2\lbd}\al_i,\ i=1,2,3,$  and the roots $\al_i,\ i=1,2,3$
are determined  by the formulae (\ref{2-40})--(\ref{2-42})
depending on the $q$ sign. In the case $c_i>0, i=1,2,3$,  this type of exact solutions
 is known in applications as two-shock waves (see, e.g., \cite{liu-fokas}).
 An example of such solution is presented on Fig.5.
 %%In the case  $c_1=c_2=c_3=1, \lbd =1$ and $\ ???$

  Consider the  generalized FN equation with the fast diffusion \be\label{52*} U_t= \ (U^{-\frac{1}{2}}U_x)_{x}+  \lambda U^{\frac{1}{2}}U_{x}+\lambda_2
 U^{\frac{1}{2}}(U^{\frac{1}{2}}-\delta)(1- U),\ \delta=\frac{\lambda_2}{2\lambda^2}. \ee
In the case $\lbd_2 \not =0$ and $\lbd_3=0$,  we immediately  obtain
$p<0$ and $q=0$, therefore
%%we arrive at  the subcase 3, which
formulae   (\ref{2-42}) and (\ref{2-44*}) give  the    solution

\be\label{2-44**}
 U=\lf[\frac{\al_2
c_2\exp({\lbd_2\over2\lbd}(\lbd\al_2^2-{3\lbd_2\over2\lbd}\al_2)t-{\lbd_2\over
2\lbd}\al_2x)+\al_3
c_3\exp({\lbd_2\over2\lbd}(\lbd\al_3^2-{3\lbd_2\over2\lbd}\al_3)t-{\lbd_2\over
2\lbd}\al_3x)}{c_1+
c_2\exp({\lbd_2\over2\lbd}(\lbd\al_2^2-{3\lbd_2\over2\lbd}\al_2)t-{\lbd_2\over
2\lbd}\al_2x)+
c_3\exp({\lbd_2\over2\lbd}(\lbd\al_3^2-{3\lbd_2\over2\lbd}\al_3)t-{\lbd_2\over
2\lbd}\al_3x)}\rg]^2.\ee

 %%Generally speaking,
 Taking  into account formula (\ref{12}) with $m=-1/2$,
  we note that solution (\ref{2-44**}) with arbitrary $c_1, c_2$ and $c_3$ is not valid in the domain
$(t,x) \in \mathbb{R^+}\times \mathbb{R}$. However, setting, for example,
$c_1=0$ we obtain travelling wave solution, which is valid  in this domain.
 An example of such solution is presented on Fig.6.
 It should be stressed that similar solutions possesses also
  the classical FN equation \cite{kawa-tana-83}
  and the generalized FN equation \cite{che-2006}
  \[ U_t = U_{xx} + \lambda UU_x+\lambda_3 U(U-\delta)(1-U),\quad
  0<\delta<1.
 \]

Subcase 4.  If $p^3+q^2>0$  then three roots of (\ref{2-29}) are different and two of them
are complex conjugate.  The   Cardano formulae should be again applied.
 Setting  $k_1=\al ,$
$k_{2,3}= a \pm i b ,$  where
\be\label{2-45}\ba{l}\al =\sqrt[3]{-q+\sqrt{p^3+q^2}}-\sqrt[3]{q+\sqrt{p^3+q^2}},\\
 a =-{1\over2}\lf(\sqrt[3]{-q+\sqrt{p^3+q^2}}-\sqrt[3]{q+\sqrt{p^3+q^2}}\rg),\\
 b
 ={\sqrt{3}\over2}\lf(\sqrt[3]{-q+\sqrt{p^3+q^2}}+\sqrt[3]{q+\sqrt{p^3+q^2}}\rg),\ea\ee
the general solution of  (\ref{2-27}) may be presented in the form
\[W=f\exp(\al  y)+\biggr(g\cos( b y)+h\sin( b y)\biggl)\exp( a  y), \quad \al=-2a.\]
 Using (\ref{2-28}) one arrives at the general solution
\[V=\frac{\al f\exp(\al  y)+\biggr(g( a \cos( b y)- b
\sin( b y))+h( b \cos( b y)+ a \sin( b y))\biggl)\exp( a
y)}{f\exp(\al  y)+\biggr(g\cos( b y)+h\sin( b y)\biggl)\exp( a
y)}\]of the non-linear ODE (\ref{2-25}). The analog of (\ref{2-44})
in this case takes the form
 \be\label{2-47}\ba{l}-3 a (f_tg-fg_t)+ b
(fh_t-f_th)= b f h(\lbd_2^*-2{\lbd_1^*}^2 a )
+\\\medskip+{\lbd_1^*\over3}\lbd a f g(2 a ^2+2 b ^2+5 a ^3+ a  b ^2),\\
-3 a (f_th-fh_t)+ b (f_tg-fg_t)=- b f g(\lbd_2^*-2{\lbd_1^*}^2 a )
+\\\medskip{\lbd_1^*\over3}\lbd a f h(2 a ^2+2 b ^2+5 a ^3+ a  b ^2),\\
gh_t-g_th={\lbd_1^*\over3} b (2\lbd a +\lbd_2)(g^2+h^2).\ea\ee
It
should be stressed that the ODE system (\ref{2-47}) has essentially
different structure from those presented above and its solving takes
a lot efforts. We were  able to realize all necessary computations,
which are omitting here,
 and to check the result using the program package MATHEMATICA 5.0.
 Finally, the exact solution
 \be\label{2-48}
U=\lf[ \frac{-2c_1 a \exp\lf(\beta_0x+\beta_1t\rg)
+c_2\biggr( a \cos\lf(\beta_2x+\beta_3t-c_3\rg)- b
\sin\lf(\beta_2x+\beta_3t-c_3\rg)\biggl)}{b_1\exp\lf(\beta_0x+\beta_1t\rg)
+b_2\cos\lf(\beta_2x+\beta_3t-c_3\rg)}\rg]^2\ee
%\[U=\lf[\frac{-2b_1 a \exp\lf(- a x+\beta_1t\rg)
%+b_2\sqrt{ a ^2+ b ^2}\cos\lf(\arccos\lf({ a \over\sqrt{ a ^2+ b ^2}}+{\lbd_1^*\over3} b x+\beta_2t-b_3\rg)\rg)}{b_1\exp\lf(- a x+\beta_1t\rg)
%+b_2\cos\lf({\lbd_1^*\over3} b x+\beta_2t-b_3\rg)}\rg]^2,\]
of the nonlinear RDC equation (\ref{52}) with $\lbd_1\not=0$
has been found. Here $\beta_0=-{3\lbd_1\over2\lbd} a ,\
\beta_1=-{\lbd_1\over2\lbd}(\lbd( b ^2+3 a ^2)-{9\lbd_1\over2\lbd} a
),\ \beta_2={\lbd_1\over2\lbd} b ,\
 \beta_3=-{\lbd_1\over2\lbd} b (2\lbd a +{3\lbd_1\over2\lbd}),$  and
$a $ and $ b$ are determined  by the formulae (\ref{2-45}).
  Note that quasi-periodic periodic solutions of
the similar form  were also  obtained for the reaction-diffusion
equation \[W_t =  W_{yy} + \lambda WW_y+\lambda_0+\lambda_1W
-\lambda_3  W^3 \] with $\lambda=0$  \cite{cla, dix-cla}.

%%\newpage

\begin{center}
{\textbf{3. Conclusions.}}
\end{center}

In the first part of this paper \cite{ch-pl-06ar}, Theorems 1 and 2
giving a complete description of $Q$-conditional symmetries of the
nonlinear RDC equations (\ref{1}) -- (\ref{1b}) are proved. It
should be stressed that all $Q$-conditional symmetry operators
listed in Theorems 1--2 \cite{ch-pl-06ar} contains the same
nonlinearities with respect to  the dependent variable $U$ as the
relevant RDC equations.
 Analogous results were earlier obtained   for
single reaction-diffusion  equations \cite{cla}, \cite{Fush93},
\cite{se-90}, \cite{nucci1}.

However, we note that there is the essential difference between RDC
equations (\ref{1}) -- (\ref{1b})  and the relevant RD equation
\[U_t=[U^mU_x]_x+C(U).\] For example, the Murray type  equation \[ U_t = U_{xx} + \lambda
UU_x+\lambda_0
%%+\lambda_1 U
+\lambda_2  U^2, \quad \lambda_2\not=0,
 \]
%% with the quadratic nonlinearity
  admits the $Q$-conditional symmetry
 \[\quad \quad \quad  Q =\partial_t +\lf(-\lambda
U+{\lambda_2 \over \lambda}\rg)\partial_x + (\lambda_0
%%+ \lambda_1U
+\lambda_2U^2 )\partial_U, \] while the RD equation with this term,
i.e. the Fisher type equation
\[U_t = U_{xx} + \lambda_0+\lambda_1 U+\lambda_2  U^2, \quad  \lambda_2\not=0
\]
 does not possess one. Similarly, the RDC equation (\ref{52}) possessing  the $Q$-conditional symmetry
 (\ref{53}) has no analog among  reaction-diffusion  equations with the diffusivity $U^{-{1 \over 2}}$.

%%It should be stressed that
The RDC equations listed in Theorems 1 and 2 \cite{ch-pl-06ar} contain
several well-known equations arising in applications and their
direct generalizations. In the particular case, the Murray equation,
 its porous  analog (\ref{3*}) and its analog  with the fast diffusion
\[U_t= \ [U^{-2}  U_{x}]_{x}+  \lambda
U^{-2}U_{x}+\lambda_1U(1- U); \]  
 the Fitzhugh-Nagumo equation \cite{fitzhugh} with the convective
term
  \[ U_t = U_{xx} + \lambda UU_x+\lambda_3 U(U-\delta)(1-U),\quad 0<\delta<1
 \]
 and its analog (\ref{52*}) with the  fast diffusion;
 the  Kolmogorov-Petrovskii-Piskunov  equation \cite{kpp-1937} with the convective term
  \[ U_t = U_{xx} + \lambda UU_x+\lambda_3 U(1-U)^2,
 \]
and  the Newell-Whitehead equation \cite{new-wh}
 with the convective term
  \[ U_t = U_{xx} + \lambda UU_x+\lambda_3 U^3- \lambda_1 U.
 \]
%%and its porous  analog, see (\ref{52}) with $ \lbd_2=0$ and $m=2$.

A further generalization of the RDC equations (\ref{1}) and (\ref{1b}) reads as
 \be \label{1-ad} U_t=[U^mU_x]_x+\lambda
U^{n}U_x+C(U), \quad \lambda \not=0, \ee
 where
 $m$ and $ n$ are   arbitrary constants.  The work is in progress
 on the complete description of $Q$-conditional symmetry of (\ref{1-ad}) and the RDC
 equation with exponential nonlinearities
 \[ U_t=[\exp(mU)U_x]_x+\lambda
\exp(nU)U_x+C(U), \quad \lambda \not=0. \]

It is well-known that new $Q$-conditional symmetries don't guarantee
the construction of  exact solutions, which cannot be obtained by
the Lie machinery (see non-trivial examples in \cite{ch-96,ch98}).
In this paper, several exact solutions were constructed using the
conditional symmetries arising in Theorem 1 and 2 \cite{ch-pl-06ar}.
It was shown that these solutions are not obtainable by Lie
symmetries, however, they contain the known plane wave solutions as
particular cases. Many of the solutions obtained possess attractive
properties and can be used for further investigation of the relevant
boundary-value problems. In the particular case, we established that
the zero Dirichlet and Neumann conditions, i.e. typical boundary
conditions for mathematical models arising in physics and biology,
can be satisfied by the relevant fitting of constants $c_1,\ c_2$
and $c_3$ (see the solutions presented on Fig.1-2, 4-6).

To our best knowledge many of the solutions presented above are new.
However, we noted that some of them can be derived from the recent
paper \cite{c-f-m}. In fact, if one applies substitution (\ref{12})
to the RDC equation (\ref{2-10}) and its solutions
(\ref{2-7})-(\ref{2-9}), then equation (54)\cite{c-f-m} with
(65)-(66) \cite{c-f-m} and $\al(s)=s^n,$ and solutions (69)-(71)
\cite{c-f-m} are exactly obtained. Nevertheless the authors of that
paper don't use any symmetries to construct exact solutions, formula
(72) \cite{c-f-m} is nothing else but the equation $Q(V)=0,$ where
$Q$ is the conditional symmetry operator \[
Q=\partial_{t}+(\lambda_1^{*} V+\lambda_2^{*})\partial_V\] of
\[ V_{xx}=V^{n}V_{t}-\lambda V_{x}+(\lambda_1^{*}
V+\lambda_2^{*})(\lambda_{3}-V^{n}).\] Thus, we obtain new
confirmation of the known idea (see, i.e. \cite{Fush93},
\cite{bl-c}) that any exact solution can be
obtained by the relevant Lie or conditional symmetry operator.

\end{document}